\documentclass[12pt]{article}
\usepackage{amsmath}
\def\a{\alpha}
\def\b{\beta}

\def\d{\delta}

\def\p{\partial}

\def\be{\begin{equation}}
\def\ee{\end{equation}}
\def\arr{\begin{array}{rll}}
\def\ea{\end{array}}
\def\bea{\begin{eqnarray}}
\def\eea{\end{eqnarray}}
\def\N2{$N{=}2$}

\def\sfrac#1#2{{\textstyle\frac{#1}{#2}}}
\def\>{\rangle}
\def\<{\langle}
\def\+{\dagger}
\def\={\ =\ }

\begin{document}
\begin{titlepage}
\setcounter{page}{0}
\begin{center}
{\Large\bf  Bianchi type-V spinning particle on $\mathcal{S}^2$}\\
%\vskip 0.4cm
%{\Large\bf   of integrable spherical mechanics }\\
\vskip 1.5cm
\textrm{\Large Anton Galajinsky \ }
\vskip 0.7cm
{\it
Tomsk Polytechnic University, 634050 Tomsk, Lenin Ave. 30, Russia} \\
\vskip 0.2cm
{\it
Tomsk State University of Control Systems and Radioelectronics, 634050 Tomsk, Lenin Ave. 40, Russia} \\
\vskip 0.2cm
{e-mail: galajin@tpu.ru}
\vskip 0.5cm
\end{center}

\begin{abstract} \noindent
Integrable spinning extension of a free particle on $\mathcal{S}^2$ is constructed in which spin degrees of freedom are represented by a $3$--vector obeying the Bianchi type--V algebra. Generalizations involving a scalar potential giving rise to two quadratic constants of the motion, or external field of the Dirac monopole, or the motion on the group manifold of $SU(2)$ are built. A link to the model of a relativistic spinning particle propagating on the near horizon $7d$ Myers--Perry black hole background is considered. Implications of the construction in this work for the $D(2,1;\alpha)$ superconformal mechanics are discussed.
\end{abstract}

\vspace{0.5cm}

\end{titlepage}
\renewcommand{\thefootnote}{\arabic{footnote}}
\setcounter{footnote}0

\noindent
{\bf 1. Introduction}\\

Near horizon geometries of extreme black holes in various dimensions attracted recently considerable attention \cite{KL}. Although the primary concern was to understand the Kerr/CFT--correspondence \cite{Com}, other aspects including the study of superconformal mechanics \cite{G1}--\cite{G3} and the construction of integrable systems \cite{G4}--\cite{DNSS} received interest.

Similarly to a generic conformal mechanics \cite{HKLN}, an integrable model associated with a near horizon black hole geometry is usually constructed by focusing on a relativistic massive spinless particle, building constants of the motion which are linked to Killing vector fields, computing the Casimir element of $SO(2,1)$ subgroup of the full isometry group, and identifying the latter with the Hamiltonian of a reduced spherical mechanics. If the isometry group involves $U(n)$ factor, one can alternatively use quadratic Casimir element of $u(n)$ so as to specify the reduced integrable system. Note that this scheme is entirely group theoretical, in which little depends on the peculiarities of a background field configuration.

A recent study of $su(2)$ spinning extensions of spherical mechanics \cite{GL} revealed a tighter connection between the original metric and the resulting integrable system. In particular, the background geometry determines a nontrivial Poisson structure on the phase space which includes spin degrees of freedom \cite{AKH}.

In Ref. \cite{GL}, the spin sector of a particle on $\mathcal{S}^2$ was represented by a symmetric Euler top. Within the Hamiltonian framework, the latter is usually described by a 3--vector obeying the structure relations of $su(2)$. Classification of three--dimensional real Lie algebras dates back to the work of Bianchi \cite{B}, in which $su(2)$ was identified with the type--IX algebra. It is natural to wonder whether other instances from the Bianchi classification (see Sect. 6) may give rise to integrable spinning particles on $\mathcal{S}^2$. The goal of this paper is to report on the Bianchi type--V model of such a kind.

The work is organized as follows. In the next section, the near horizon $7d$ Myers--Perry black hole metric is given in spherical coordinates and the full group of isometry transformations is exposed. In Sect. 3 a relativistic massive spinning particle on the curved background is considered within the formalism introduced in Ref. \cite{AKH} and possible consistent truncations are discussed which may result in integrable spinning extensions of a particle on $\mathcal{S}^2$. In Sect. 4 we focus on the case in which spin degrees of freedom are represented by a $3$--vector satisfying the Bianchi type--V algebra. The dynamics of the system is studied in detail and its minimal superintegrability is established. Three generalizations of the Bianchi type--V spinning particle on $\mathcal{S}^2$, which maintain minimal superintegrability, are discussed next. The first of them involves an extra scalar potential which gives rise to two quadratic constants of the motion. It can be regarded as the Bianchi type--V spinning extension of the case $(S9)$ in Ref. \cite{KKPM}. The second model takes into account external field of the Dirac monopole. The third system describes a Bianchi type--V spinning particle moving on the group manifold of $SU(2)$. In Sect. 5 it is demonstrated that each realization of $su(2)$ in Sect. 4 can be used to generate a novel example of the $D(2,1;\alpha)$ superconformal mechanics. The concluding Sect. 6 contains a brief account of how the consideration in this work could be extended to cover other instances from the Bianchi classification.

\vspace{0.5cm}

\noindent
{\bf 2. Background metric and its symmetries}\\

Our starting point is the
near horizon $7d$ Myers--Perry black hole metric in which all rotation parameters are set equal\footnote{Here and in what follows we use the notations in \cite{GNS}. Constant factors entering the metric were chosen so as to keep the Killing vector fields (\ref{u3}) in the simplest form. }
\bea\label{metric}
&&
ds^2=-r^2 dt^2+\frac{dr^2}{r^2}+12 \left(d\theta^2+\sin^2{\theta} d\phi^2\right)+2 \sin^2{\phi} \sin^2{\theta}{\left(r dt + 3 \sqrt{2} d\psi_1\right)}^2
\nonumber\\[2pt]
&&
\qquad \quad
+2 \cos^2{\phi} \sin^2{\theta} {\left(r dt + 3 \sqrt{2} d\psi_2\right)}^2+2 \cos^2{\theta} {\left(r dt + 3 \sqrt{2} d\psi_3\right)}^2
\nonumber\\[2pt]
&&
\qquad \quad -24 \cos^2{\phi}\sin^2{\phi}\sin^4{\theta}{\left(d\psi_1-d\psi_2 \right)}^2-24\sin^2{\phi}\cos^2{\theta}\sin^2{\theta}{\left(d\psi_1-d\psi_3 \right)}^2
\nonumber\\[2pt]
&&
\qquad \quad
-24\cos^2{\phi}\cos^2{\theta}\sin^2{\theta}{\left(d\psi_2-d\psi_3 \right)}^2.
\eea
Here $t$ and $r$ are the temporal and radial variables, $(\theta,\phi)$ are the latitudinal angular coordinates, and
$(\psi_1,\psi_2,\psi_3)$ are the azimuthal angles. It is straightforward to verify that (\ref{metric}) is a particular solution of
the vacuum Einstein equations.

A key ingredient of the construction to follow is to convert abundant symmetries of the metric (\ref{metric}) into constants of the motion which characterize an integrable system. It is thus important to have a clear understanding of Killing vectors associated with (\ref{metric}).

In addition to the Killing vector fields
\bea
&&
H=\partial_t, \quad D=t \partial_t-r \partial_r, \quad K=\left(t^2+\frac{1}{r^2}\right)\partial_t-2 t r \partial_r-\frac{\sqrt{2}}{3 r   }\left(\partial_{\psi_1}+\partial_{\psi_2}+\partial_{\psi_3}\right),
\eea
which form the conformal algebra $so(2,1)$ and are common for most of the near horizon black hole geometries \cite{KL},
the metric (\ref{metric}) enjoys $U(3)$--symmetry generated by the operators $\xi_{ij}=-\xi_{ji}$, $\rho_{ij}=\rho_{ji}$ which read
\bea\label{u3}
&&
\xi_{12}=- \cos{\psi_{12}} \p_{\phi} +\sin{\psi_{12}}\left(\cot{\phi} \p_{\psi_1}  +\tan{\phi} \p_{\psi_2} \right) ,
\nonumber\\[4pt]
&&
\xi_{13}=- \cos{\psi_{13}} \left(\sin{\phi}\p_{\theta}+\cos{\phi} \cot{\theta} \p_{\phi} \right) +
\sin{\psi_{13}}\left( \frac{\cot{\theta}}{\sin{\phi}}\p_{\psi_1}+ \sin{\phi} \tan{\theta}\p_{\psi_3}\right) ,
\nonumber\\[4pt]
&&
\xi_{23}=\cos{\psi_{23}}\left( - \cos{\phi}\p_{\theta}+\sin{\phi} \cot{\theta} \p_{\phi}\right) +
\sin{\psi_{23}}\left( \frac{\cot{\theta}}{\cos{\phi}}\p_{\psi_2}+ \cos{\phi} \tan{\theta}\p_{\psi_3}\right) ,
\nonumber\\[4pt]
&&
\rho_{12}=\sin{\psi_{12}} \p_{\phi}  +\cos{\psi_{12}} \left( \cot{\phi}\p_{\psi_1} +\tan{\phi}\p_{\psi_2} \right) ,
\nonumber\\[4pt]
&&
\rho_{13}=\sin{\psi_{13}} \left(\sin{\phi}\p_{\theta}+\cos{\phi} \cot{\theta} \p_{\phi}\right)  +
\cos{\psi_{13}} \left( \frac{\cot{\theta}}{\sin{\phi}}\p_{\psi_1}+ \sin{\phi} \tan{\theta}\p_{\psi_3}\right),
\nonumber\\[4pt]
&&
\rho_{23}=\sin{\psi_{23}} \left(\cos{\phi}\p_{\theta}- \sin{\phi} \cot{\theta} \p_{\phi} \right) +
\cos{\psi_{23}} \left( \frac{\cot{\theta}}{\cos{\phi}}\p_{\psi_2}+ \cos{\phi} \tan{\theta}\p_{\psi_3}\right) ,
\nonumber\\[4pt]
&&
\rho_{11}=2\p_{\psi_1}, \qquad \rho_{22}=2\p_{\psi_2}, \qquad \rho_{33}=2\p_{\psi_3},
\eea
where we abbreviated $\psi_{ij}=\psi_i-\psi_j$, $i,j=1,2,3$. Note that
the enhancement of $U(1)^3$ symmetry associated with the transformations $\psi'_i=\psi_i+\lambda_i$ to a larger group $U(3)$ is a consequence of setting all rotation parameters to be equal to each other.

\vspace{0.5cm}

\noindent
{\bf 3. Spinning particle on a curved background and reduced spherical mechanics}\\

Our strategy in constructing spinning extensions of integrable spherical mechanics is to implement a proper reduction of the model of a relativistic massive spinning particle propagating on the curved background with $SU(3)$ isometry group. To this end, let us consider a phase space parametrized by the canonical pair $(x^\mu,p_\mu)$ and the self-conjugate spin variables $S^{\mu\nu}=-S^{\nu\mu}$, $\mu,\nu=0,\dots,6$, which is endowed with
the Poisson structure introduced in \cite{AKH}
\bea\label{br}
&&
\{x^\mu,p_\nu \}={\delta^\mu}_\nu, \quad \{p_\mu,p_\nu\}=-\sfrac 12 R_{\mu\nu\lambda\sigma} S^{\lambda\sigma}, \quad \{S^{\mu\nu},p_\lambda \}=\Gamma^\mu_{\lambda\sigma} S^{\nu\sigma}-\Gamma^\nu_{\lambda\sigma} S^{\mu\sigma},
\nonumber\\[2pt]
&&
\{S^{\mu\nu},S^{\lambda\sigma} \}=g^{\mu\lambda} S^{\nu\sigma}+g^{\nu\sigma} S^{\mu\lambda}-g^{\mu\sigma} S^{\nu\lambda}-g^{\nu\lambda} S^{\mu\sigma}.
\eea
Here $g^{\mu\nu}$ is the inverse metric, $\Gamma^\mu_{\lambda\sigma}$ are the Christoffel symbols, and
$R_{\mu\nu\lambda\sigma}$ is the Riemann tensor. Within this framework, to each Killing vector field $\lambda^\mu (x) \p_\mu$ characterizing the background geometry there corresponds the first integral of the relativistic equations of motion \cite{AKH}
\be\label{Int}
\tilde\lambda=\lambda^\mu p_\mu+\sfrac12 \nabla_\mu \lambda_\nu S^{\mu\nu}.
\ee
For what follows, it is important to emphasize  that the algebra formed by the phase space functions (\ref{Int}) under the bracket (\ref{br}) is analogous to that generated by Killing vector fields.

Taking into account the recent studies in \cite{GL}, a reasonable truncation of the model would be to abandon the temporal and radial coordinates along with the Mathisson–-Papapetrou–-Dixon equations and to focus on the angular sector whose dynamics is governed by quadratic Casimir invariant of $su(3)$. Spin degrees of freedom which do not contribute into $su(3)$ generators are to be discarded as well. Unfortunately, the resulting Hamiltonian and $su(3)$ conserved charges turn out to be extremely bulky.

For lack of a better alternative, we are led to further simplify the model by implementing the reduction proposed in \cite{GNS}. Taking into account that the azimuthal angular variables $(\psi_1,\psi_2,\psi_3)$ are cyclic, it is natural to set the integrals of motion (\ref{Int}) associated with the Killing vector fields $\rho_{11}$, $\rho_{22}$, $\rho_{33}$ in (\ref{u3}) to be equal to (coupling) constants $g_1$, $g_2$, and $g_3$, respectively. Analysing the structure relations of $su(3)$,
one can then verify that the quadratic combinations
\be\label{I}
I_{ij}={\tilde\xi}_{ij}^2+{\tilde\rho}_{ij}^2, \qquad i<j,
\ee
where ${\tilde\xi}_{ij}$ and ${\tilde\rho}_{ij}$ denote the phase space functions (\ref{Int}) associated with the Killing vector fields (\ref{u3}),
commute with $(\tilde\rho_{11},\tilde\rho_{22},\tilde\rho_{33})$, and represent the integrals of motion of a dynamical system governed by the Hamiltonian
\be\label{HHH}
H=\sum_{i<j} I_{ij}.
\ee
The explicit expressions for $I_{ij}$ read
\bea\label{III}
&&
I_{12}=\left(p_{\phi}+\sin{2\phi} \sin^2{\theta} \left(S^{t\psi_1} - S^{t\psi_2}\right)-6\sin{2\phi} \sin^2{\theta}(2-\cos{2\theta}) S^{\psi_1 \psi_2}
\right.
\nonumber\\[2pt]
&&
\qquad ~
{\left.
+6\sin{2\theta} S^{\theta\phi}-3 \sin{2\phi}\sin^2{2\theta}\left(S^{\psi_1 \psi_3}-S^{\psi_2 \psi_3}\right)\right)}^2
\nonumber\\[2pt]
&&
\qquad ~
+{\left(g_1 \cot{\phi}+g_2 \tan{\phi} -12  \sin^2{\theta} \left(S^{\phi \psi_1 }-S^{\phi \psi_2}\right)\right)}^2,
\nonumber\\[4pt]
&&
I_{13}=\left(p_{\phi} \cos{\phi} \cot{\theta}+p_{\theta} \sin{\phi}+\sin{\phi}\sin{2\theta} \left(S^{t\psi_1} - S^{t\psi_3}\right)-12 \cos{\phi} \sin^2{\theta} S^{\theta\phi }
\right.
\nonumber\\[2pt]
&&
\qquad ~
-24 \sin{\phi} \cos^2{\phi}\cos{\theta} \sin^3{\theta} (S^{\psi_1\psi_2}+S^{\psi_2\psi_3})
\nonumber\\[2pt]
&&
\qquad ~
{\left.
-3\sin{\phi} (\cos^2{\phi} \sin{4\theta}+(5-\cos{2\phi})\sin{2\theta})S^{\psi_1\psi_3}
\right)}^2
\nonumber\\[2pt]
&&
\qquad ~
+\left(g_1 \sin^{-1}{\phi}\cot{\theta}+g_3 \sin{\phi} \tan{\theta} -6 \cos{\phi} \sin{2\theta} \left(S^{\phi \psi_1 }-S^{\phi \psi_3}\right)
\right.
\nonumber\\[2pt]
&&
\qquad ~
{\left.
-12\sin{\phi}\left(S^{\theta \psi_1 }-S^{\theta \psi_3}\right) \right)}^2,
\nonumber\\[4pt]
&&
I_{23}=\left(p_{\phi} \sin{\phi} \cot{\theta}-p_{\theta} \cos{\phi}
-\cos{\phi}\sin{2\theta}\left(S^{t\psi_2} - S^{t\psi_3}\right)-12 \sin{\phi} \sin^2{\theta} S^{ \theta\phi}
\right.
\nonumber\\[2pt]
&&
\qquad ~
-24\cos{\phi}\sin^2{\phi}\cos{\theta}\sin^3{\theta} \left(S^{\psi_1\psi_2} - S^{\psi_1\psi_3}\right)
\nonumber\\[2pt]
&&
\qquad ~
{\left.
+\frac 32 \left((11 \cos{\phi}+\cos{3\phi})\sin{2\theta}+\sin{\phi}\sin{2\phi}\sin{4\theta}\right)S^{\psi_2\psi_3}
\right)}^2
\nonumber\\[2pt]
&&
\qquad ~
+\left(g_2 \cos^{-1}{\phi}\cot{\theta}+g_3 \cos{\phi} \tan{\theta} +6 \sin{\phi} \sin{2\theta} \left(S^{\phi \psi_2 }-S^{\phi \psi_3}\right)
\right.
\nonumber\\[2pt]
&&
\qquad ~
{\left.
-12\cos{\phi}\left(S^{\theta \psi_2 }-S^{\theta \psi_3}\right) \right)}^2.
\eea
In order to simplify these formulae and the algebra (\ref{BRR}) below, we fixed the value of the radial coordinate at $r=\frac{1}{3\sqrt{2}}$. As is customary in the literature on relativistic spinning particles propagating on curved backgrounds, the constituents of the spin tensor $S^{\mu\nu}$ are labeled by the same letters which designate the coordinates. The explicit form of the Poisson structure relations (\ref{br}) originating from the metric (\ref{metric}) is bulky and will not be exposed here.

The resulting model still looks rather complicated. Furthermore, the thirteen spin degrees of freedom entering $H$ leave little hope that it is actually integrable.
A natural way out is to get rid of some of the spin variables. A rigorous procedure of doing so is to consider an even number of spin degrees of freedom, which have a non--degenerate bracket among themselves, regard half of them as first class constrains and the other half as gauge fixing conditions and discard the full set after introducing the Dirac bracket.

\vspace{0.5cm}

\noindent
{\bf 4. Bianchi type-V spinning extension of integrable spherical mechanics}\\

Searching for a nontrivial yet tractable integrable system originating from Eqs. (\ref{HHH}), (\ref{III}), in this section we consider the case in which ten spin degrees of freedom are eliminated from the consideration while the rest obeys the Bianchi type-V algebra.

Let us choose $S^{t\psi_1}:=-J_1$, $S^{t\psi_2}:=-J_2$, $S^{t\psi_3}:=-J_3$ to be the spin variables, which survive the reduction, and impose the Poisson structure relations
\bea\label{BRR}
\{\theta,p_{\theta}\}=1, \qquad \{\phi,p_{\phi}\}=1, \qquad \{J_i,J_j\}=-J_i+J_j,
\eea
with $i,j=1,2,3$, for which the Jacobi identities are automatically satisfied. Eqs. (\ref{III}) give rise to the phase space functions
\bea\label{II}
&&
I_{12}={\left(p_{\phi}-(J_1-J_2)\sin{2\phi}\sin^2{\theta} \right)}^2+{\left(g_1 \cot{\phi}+g_2 \tan{\phi} \right)}^2,
\nonumber\\[4pt]
&&
I_{13}={\left(-p_{\phi} \cos{\phi} \cot{\theta}-p_{\theta}\sin{\phi}+(J_1-J_3)\sin{\phi}\sin{2\theta} \right)}^2
\nonumber\\[2pt]
&&
\qquad ~
+{\left(g_1 \sin^{-1}{\phi}\cot{\theta}+g_3 \sin{\phi} \tan{\theta}\right)}^2,
\nonumber\\[2pt]
&&
I_{23}={\left(-p_{\phi} \sin{\phi} \cot{\theta}+p_{\theta}\cos{\phi}-(J_2-J_3)\cos{\phi}\sin{2\theta} \right)}^2
\nonumber\\[4pt]
&&
\qquad ~
+{\left(g_2 \cos^{-1}{\phi}\cot{\theta}+g_3 \cos{\phi} \tan{\theta}\right)}^2,
\eea
which commute with the Hamiltonian
\be\label{HH}
H=\frac 12 \sum_{i<j} I_{ij},
\ee
under the bracket (\ref{BRR}). Taking into account that
\be\label{com}
J_1-J_2, \qquad J_1-J_3
\ee
are conserved as well, one concludes that (\ref{HH}) defines an integrable system. Four functionally independent integrals of motion in involution include $(H,I_{12},J_1-J_2,J_1-J_3)$, while the extra constant of the motion $I_{13}$ (or alternatively $I_{23}$) ensures minimal superintegrability.
Omitting the spin degrees of freedom $(J_1,J_2,J_3)$, one reveals a particle on $\mathcal{S}^2$ in the presence of an external scalar potential, which admits
two quadratic constants of the motion. In the terminology of Ref. \cite{KKPM} it corresponds to the case $(S9)$.

Setting the coupling constants $(g_1,g_2,g_3)$ to vanish, one gets the Bianchi type-V spinning extension of a free particle on $\mathcal{S}^2$. Indeed, it is straightforward to verify that the phase space functions
\bea\label{BTV}
&&
\mathcal{J}_1=-p_{\phi} \cos{\phi} \cot{\theta}-p_{\theta}\sin{\phi}+(J_1-J_3)\sin{\phi}\sin{2\theta},
\nonumber\\[2pt]
&&
\mathcal{J}_2=-p_{\phi} \sin{\phi} \cot{\theta}+p_{\theta}\cos{\phi}-(J_2-J_3)\cos{\phi}\sin{2\theta},
\nonumber\\[2pt]
&&
\mathcal{J}_3=p_{\phi}-(J_1-J_2)\sin{2\phi}\sin^2{\theta},
\eea
which derive from $\sqrt{I_{ij}}$ at $g_1=g_2=g_3=0$,
obey the structure relations of $su(2)$ under the bracket (\ref{BRR})
\be
\{\mathcal{J}_i,\mathcal{J}_j\}=\epsilon_{ijk} \mathcal{J}_k,
\ee
where $\epsilon_{ijk}$ is the Levi--Civita symbol with $\epsilon_{123}=1$. The Hamiltonian (\ref{HH}) reduces to the Casimir element
\be\label{Ham}
H=\frac 12  \vec{\mathcal{J}}^2.
\ee
Four functionally independent integrals of motion in involution $(H,\mathcal{J}_3,J_1-J_2,J_1-J_3)$ provide the Liouville integrability. One more constant of the motion $\mathcal{J}_1$ (or alternatively $\mathcal{J}_2$) renders the model minimally superintegrable.

Let us discuss qualitative dynamical behaviour of the system.
Turning to the Cartesian coordinates
\be\label{sphcoord}
\vec{x}=(\cos{\phi}\sin{\theta},\sin{\phi}\sin{\theta},\cos{\theta}),
\ee
and analysing Hamilton's equations following from (\ref{BRR}), (\ref{Ham}), one reveals the uniform motion on $\mathcal{S}^2$
\be\label{cone}
\dot{\vec{x}}^2=\dot\theta^2+\dot\phi^2 \sin^2{\theta}=2H=\textrm{const}
\ee
which takes place along the intersection of the sphere and the plane passing through the origin
\be
\vec{x} \cdot \vec{\mathcal{J}}=0,
\ee
i.e. the great circle. Note that $H$ entering (\ref{cone}) is the energy of the {\it full} system. It is here that one sees the impact of the spin degrees of freedom upon the orbital motion on $\mathcal{S}^2$.

Taking into account the integrals of motion (\ref{com}) and introducing the constant vector $\vec{m}=(J_2-J_3,J_3-J_1,J_1-J_2)$, one similarly finds that evolution of the spin degrees of freedom is confined to the plane in $3d$ spin subspace
\be\label{plane}
\vec{J} \cdot \vec{m}=0.
\ee
Hamilton's equation resulting from (\ref{BRR}), (\ref{Ham}) and specifying the change of $\vec{J}$ over time reads
\bea\label{efm}
&&
\dot{\vec{J}}=\omega(t)\vec{l},
\eea
where $\vec{l}=(1,1,1)$ and
\bea
&&
\omega(t)=\mathcal{J}_1 (J_1-J_3) \sin{\phi(t)} \sin{2\theta(t)}-\mathcal{J}_2 (J_2-J_3)\cos{\phi(t)} \sin{2\theta(t)}
\nonumber\\[2pt]
&&
\qquad ~
-\mathcal{J}_3 (J_1-J_2) \sin{2\phi(t)} \sin^2{\theta(t)}.
\eea
It has the general solution
\be
\vec{J}=\vec{\alpha}+\vec{l} \int_{t_0}^t d\tau \omega(\tau),
\ee
where $\vec{\alpha}$ is an arbitrary vector belonging to the plane (\ref{plane}), i.e. a constant of integration obeying $\vec{\alpha} \cdot \vec{m}=0$.

Thus, while a particle on $\mathcal{S}^2$ is travelling along the great circle, the tip of the spin vector $\vec{J}(t)$ oscillates forwards and backwards along the straight line parallel to $\vec{l}=(1,1,1)$ in $3d$ spin subspace. This is in agreement with geometric interpretation of constants of the motion (\ref{com}) as two planes
\bea
&&
\vec{J} \cdot \vec{n}_1=\textrm{const}, \qquad \vec{J} \cdot \vec{n}_2=\textrm{const}',
\eea
with the normals $\vec{n}_1=(1,-1,0)$ and $\vec{n}_2=(1,0,-1)$, respectively, which intersect along the straight line parallel to $\vec{l}=\vec{n}_1 \times \vec{n}_2$. The swinging of $\vec{J}$ looks most transparent for a particular solution corresponding to the orbital motion in the equatorial plane ($\mathcal{J}_1=\mathcal{J}_2=p_{\theta}=0$)
\be
\theta=\frac{\pi}{2}, \qquad \phi(t)=\mathcal{J}_3 t+\beta, \qquad \vec{J}(t)=\vec{\alpha}+\frac 12 \vec{l}  (J_1-J_2) \cos{2\phi(t)},
\ee
where $\beta$ and $\vec{\alpha}$ are constants of integration.

It is interesting to compare the model above with the Bianchi type--IX spinning particle on $\mathcal{S}^2$ constructed recently in \cite{GL}. In the latter case, the $J_2$--component of the spin vector $\vec{J}$ is conserved,
a particle parametrized by the angular variables $(\theta,\phi)$ follows a circular orbit, which is the intersection of $\mathcal{S}^2$ and the cone
\be
x_i \mathcal{J}_i=-J_2=\textrm{const},
\ee
while the spin vector precesses around $J_2$--axis in $3d$ spin subspace
\bea
&&
J_1(t)=R \cos{\Omega(t)}, \qquad J_3(t)=R\sin{\Omega(t)},
\nonumber\\[2pt]
&&
\Omega(t)=\Omega_0-\int_{t_0}^t d \tau \frac{J_2+\mathcal{J}_3\cos{\theta(\tau)}}{\sin^2 {\theta(\tau)}},
\eea
where $R$ and $\Omega_0$ are constants of integration.

Concluding this section, we note that the $su(2)$ generators in Eq. (\ref{BTV}) can be extended to include a magnetic charge $q$
\be\label{su22}
\mathcal{J}'_1=\mathcal{J}_1+q \frac{\cos{\phi}}{\sin{\theta}}, \qquad
\mathcal{J}'_2 = \mathcal{J}_2+q \frac{\sin{\phi}}{\sin{\theta}}, \qquad
\mathcal{J}'_3=\mathcal{J}_3.
\ee
Because the angular and spin degrees of freedom commute, $\mathcal{J}'_i$ obey the $su(2)$ algebra. Identifying the Casimir element with the Hamiltonian, $H'=\frac 12  \vec{\mathcal{J}'}^2$, one gets the Bianchi type-V spinning particle on $\mathcal{S}^2$ coupled to external field of the Dirac monopole.
In contrast to the previous case, a particle moves along a circular orbit, which is the intersection of $\mathcal{S}^2$ and the cone
\be
\vec{x} \cdot \vec{\mathcal{J}}'=q.
\ee
As above, the functionally independent integrals of motion in involution $(H',\mathcal{J}'_3,J_1-J_2,J_1-J_3)$ ensure the Liouville integrability, while the conservation of $\mathcal{J}'_1$ (or $\mathcal{J}'_2$) over time provides minimal superintegrability.

Finally, because $q$ in (\ref{su22}) is constant, one can introduce into the consideration an extra canonical pair $(\chi,p_\chi)$, obeying the standard bracket $\{\chi,p_\chi \}=1$, and implement the oxidation with respect to $q$
\be
q  ~ \rightarrow ~ p_\chi.
\ee
The corresponding $su(2)$ generators
\be\label{su23}
\mathcal{J}''_1=\mathcal{J}_1+\frac{\cos{\phi}}{\sin{\theta}} p_\chi , \qquad
\mathcal{J}''_2 = \mathcal{J}_2+\frac{\sin{\phi}}{\sin{\theta}} p_\chi , \qquad
\mathcal{J}''_3=\mathcal{J}_3,
\ee
and the Hamiltonian $H''=\frac 12  \vec{\mathcal{J}''}^2$ describe the Bianchi type-V spinning particle propagating on the group manifold of $SU(2)$. Its Liouville integrability is provided by five
functionally independent integrals of motion in involution $(H'',\mathcal{J}''_3,J_1-J_2,J_1-J_3,p_\chi)$. The fact that $\mathcal{J}''_1$ does not change as the system evolves over time implies the model is minimally superintegrable.

\vspace{0.5cm}

\noindent
{\bf 5. $D(2,1;\alpha)$ superconformal extensions}\\

As was demonstrated in \cite{G1}, any realization of $su(2)$ in terms of phase space functions $\mathcal{J}_i$, $\{\mathcal{J}_i,\mathcal{J}_j \}=\epsilon_{ijk}\mathcal{J}_k$, gives rise to a variant of the $D(2,1;\alpha)$ superconformal mechanics. It suffices to
introduce into the consideration an extra bosonic canonical pair $(x,p)$ along with the fermionic $SU(2)$--spinor partners\footnote{For our spinor conventions see \cite{G3}.} $(\psi_\alpha,\bar\psi^\a)$, ${(\psi_\alpha)}^{*}=\bar\psi^\a$, $\alpha=1,2$, and to impose the structure relations
\be\label{cr}
\{x,p\}=1, \qquad \{ \psi_\alpha, \bar\psi^\beta \}=-i{\delta_\a}^\b.
\ee
It is assumed that $\mathcal{J}_i$ commute with $(x,p,\psi_\alpha)$.
The generators forming Lie superalgebra associated with the exceptional supergroup $D(2,1;\alpha)$ read
\begin{align}\label{rep}
&
H=\frac{p^2}{2}+\frac{2 \alpha^2 }{x^2} \mathcal{J}_i \mathcal{J}_i+\frac{2 \alpha}{x^2} (\bar\psi \sigma_i \psi) \mathcal{J}_i -\frac{(1+2\alpha)}{4x^2} \psi^2 \bar\psi^2, && D=tH-\frac 12 x p,
\nonumber\\[2pt]
&
K=t^2 H-t x p +\frac 12 x^2, && \mathcal{I}_i=\mathcal{J}_i+\frac 12 (\bar\psi \sigma_i \psi),
\nonumber\\[2pt]
&
Q_\a=p \psi_\a-\frac{2i\alpha}{x} {(\sigma_i \psi)}_\alpha \mathcal{J}_i -\frac{i(1+2\alpha)}{2x} \bar\psi_\alpha \psi^2, && S_\alpha=x \psi_\alpha -t Q_\alpha,
\nonumber\\[2pt]
&
\bar Q^\alpha =p \bar\psi^\a+\frac{2i\alpha}{x} {(\bar\psi \sigma_i)}^\alpha \mathcal{J}_i -\frac{i(1+2\alpha)}{2x} \psi^\a \bar\psi^2, &&
\bar S^\alpha=x \bar\psi^\alpha -t \bar Q^\alpha,
\nonumber\\[2pt]
&
I_{-}=\frac{i}{2} \psi^2, \qquad \qquad \qquad \qquad I_{+}=-\frac{i}{2} {\bar\psi}^2, &&
I_3=\frac 12 \bar\psi \psi,
\end{align}
where $\sigma_i$ are the Pauli matrices. $H$ is linked to temporal translations (the Hamiltonian). $D$ and $K$ generate dilatations and special conformal transformations. $Q_\a$ are the supersymmetry generators and $S_\a$ are their superconformal partners. $\mathcal{I}_a$ form the $R$--symmetry subalgebra $su(2)$. $I_{\pm}$, $I_3$ generate one more $su(2)$.

Combining the algebraic framework above with the results in the preceding section, one concludes that each realization of $su(2)$ in Sect. 4 gives rise to a novel example of the $D(2,1;\alpha)$ superconformal mechanics. Similar models incorporating the Bianchi type-IX algebra have been studied in \cite{FIL,GL}.

If desirable, one can further generalize the system by introducing an extra fermionic canonical pair $(\chi_\alpha,\bar\chi^\a)$,
$\bar\chi^\a={(\chi_\alpha)}^{*}$, $\{ \chi_\a, \bar\chi^\b \}=-i{\d_\a}^\b$, $\alpha,\beta=1,2$, and extending the $su(2)$ generators as follows:
\be\label{s4}
\mathcal{J}_i \quad \rightarrow \quad \mathcal{J}_i+\frac 12 (\bar\chi \sigma_i \chi).
\ee
The resulting models will include an extra on--shell supermultiplet of the type $(0,4,4)$ \cite{G3}.

\vspace{0.5cm}

\noindent
{\bf 6. Discussion}\\

Three--dimensional real Lie algebras were classified in \cite{B}. Representing the structure constants in the form\footnote{We follow a modern exposition in \cite{DNF}.}
\be\label{cc}
c_{ij}^k=\epsilon_{ijl} b^{lk}-\delta_i^k a_j+\delta_j^k a_i,
\ee
where $b^{lk}$ is a symmetric $3\times 3$ matrix and $a_i$ is an arbitrary $3$--vector, and analysing the Jacobi identities, one gets the restriction
\be
b^{ij}a_j=0.
\ee
Rotating the basis in the algebra, one can diagonalize the matrix $b^{ij}=b_i \delta^{ij}$ (no sum) and bring the vector $a_i$ to the form $a_i=(a,0,0)$.
Further rescaling/relabeling of the generators reveals eleven distinct cases exposed in Table 1 above.

\begin{table}
\caption{The Bianchi classification of three--dimensional real Lie algebras}\label{table}
\begin{eqnarray*}
\begin{array}{|l|r|r|r|r|r|r|r|r|c|}
\hline
\mbox{Type} & a & b_1 & b_2 & b_3 & \mbox{Type} & a & b_1 & b_2 & b_3\\
\hline
$I$ & 0 & 0 & 0 & 0 & $V$ & 1 & 0 & 0 & 0\\
\hline
$II$ & 0 & 1 & 0 & 0 & $IV$ & 1 & 0 & 0 & 1\\
\hline
{VII}_0 & 0 & 1 & 1 & 0 & $VII$ & a & 0 & 1 & 1\\
\hline
{VI}_0 & 0 & 1 & -1 & 0 & $III$  & 1 & 0 & 1 & -1\\
\hline
$IX$ & 0 & 1 & 1 & 1 & $VI$ & a & 0 & 1 & -1\\
\hline
$VIII$ & 0 & 1 & 1 & -1 &  &   &  &  & \\
\hline
\end{array}
\end{eqnarray*}
\end{table}

The Bianchi type-V spinning particle on $\mathcal{S}^2$ constructed above\footnote{Note that in Sect. 4 a slightly different basis was chosen in which $b^{ij}=0$, $a_i=(1,1,1)$. It is related to that in Table 1 by a rotation of the basis in the algebra.} and its type-IX counterpart in \cite{GL} suggest that other instances from the Bianchi classification might give rise to potentially interesting
integrable systems of a similar kind. A reasonable strategy of building such models, is to start with the ansatz for the Poisson structure
\begin{align}\label{BR}
&
\{\theta,p_\theta\}=1, && \{\phi,p_\phi\}=1, && \{p_\theta,p_\phi\}=\alpha_i(\theta,\phi) J_i,
\nonumber\\[4pt]
&
\{J_i,p_\phi\}=\beta_{ij}(\theta,\phi) J_j, && \{J_i,p_\theta\}=\gamma_{ij}(\theta,\phi) J_j, && \{J_i,J_j \}=c_{ij}^k J_k,
\end{align}
where $\alpha_i(\theta,\phi),\beta_{ij}(\theta,\phi),\gamma_{ij}(\theta,\phi)$ are unknown functions and $c_{ij}^k$ are specified by Eq. (\ref{cc}) and the Table 1,
to extend the standard expressions for the angular momentum of a free particle on $\mathcal{S}^2$ by contributions linear in the spin variables $J_i$
\bea\label{JJJ}
&&
\mathcal{J}_1=-p_{\phi} \cos{\phi} \cot{\theta}-p_{\theta}\sin{\phi}+\sigma_i(\theta,\phi) J_i,
\nonumber\\[2pt]
&&
\mathcal{J}_2=-p_{\phi} \sin{\phi} \cot{\theta}+p_{\theta}\cos{\phi}+\mu_i(\theta,\phi) J_i,
\nonumber\\[2pt]
&&
\mathcal{J}_3=p_{\phi}+\nu_i(\theta,\phi) J_i,
\eea
where $\sigma_i(\theta,\phi),\mu_i(\theta,\phi),\nu_i(\theta,\phi)$ are unknown functions,
and to fix all the unknown coefficients from the requirements that (\ref{BR}) is compatible with the Jacobi identities, while (\ref{JJJ}) satisfy the structure relations of $su(2)$ under the bracket (\ref{BR}). These conditions will yield a coupled system of nonlinear partial differential equations which is to be solved. Similarly to the type--V and type--IX cases, it is plausible to assume that the variables $\theta$ and $\phi$ are separated, in which case the Fourier analysis might be helpful. Other consistent truncations of Eqs. (\ref{HHH}), (\ref{III}) are worth studying as well. Finally, it would be interesting to explore whether $J_i$ in Sect. 4 could be represented in terms of the Euler--like angles.

\vspace{0.5cm}

\noindent{\bf Acknowledgements}\\

\noindent
This work was supported by the Russian Science Foundation, grant No 19-11-00005.

\end{document}